\documentclass[aps,print,twocolumn,hidepacs]{revtex4}%
\usepackage{amssymb}
\usepackage{amsfonts}
\usepackage{amsmath}
\usepackage{graphicx}%
\providecommand{\U}[1]{\protect\rule{.1in}{.1in}}

\begin{document}
\title{Atomically thin spherical shell-shaped superscatterers based on Bohr model}
\author{Rujiang Li$^{1,2,3}$, Xiao Lin$^{1,2,3}$, Shisheng Lin$^{1,2}$\footnote[1]{shishenglin@zju.edu.cn},
Xu Liu$^{1}$, Hongsheng Chen$^{1,2,3}$\footnote[2]{hansomchen@zju.edu.cn}}
\affiliation{$^{1}$State Key Laboratory of Modern Optical Instrumentation, Zhejiang
University, Hangzhou 310027, China}
\affiliation{$^{2}$College of Information Science and Electronic Engineering, Zhejiang
University, Hangzhou 310027, China}
\affiliation{$^{3}$The Electromagnetics Academy of Zhejiang University, Zhejiang
University, Hangzhou 310027, China}


\begin{abstract}
Graphene monolayers
can be used for atomically thin three-dimensional shell-shaped superscatterer designs.
Due to the excitation of the first-order resonance of transverse magnetic (TM) graphene plasmons,
the scattering cross section of the bare subwavelength dielectric particle is
enhanced significantly by five orders of magnitude.
The superscattering phenomenon can be intuitively understood and interpreted with
Bohr model. Besides, based on the analysis of Bohr model, it is shown that contrary to the TM case,
superscattering is hard to occur by exciting the resonance of transverse electric (TE)
graphene plasmons due to their poor field confinements.
\end{abstract}
\maketitle

\section{Introduction}
\noindent With the concept of metamaterials and metasurfaces, subwavelength
structures can demonstrate unusual electromagnetic properties \cite%
{nmat11-917,LPR9-195},
e.g. surface plasmons in coaxial metamaterial cables \cite{JOSAB27-148,MPLB27-1330013}
and graphene-based cloaks \cite{acsnano5-5855,OE21-12592,JP27-185304}.
Specially, the scattering of subwavelength structures can be
enhanced to realize superscatterers, which can magnify the scattering cross
section of a given object remarkably \cite{OE16-18545}. Due to their
potential applications in detection, spectroscopy, and photovoltaics \cite%
{PNAS,JPCC,APL73-3815,NL8-3983,prl90-057401,JAP101-093105,nmat9-205},
various superscatterers with different shapes and types have been designed
\cite%
{NJP10-113016,APL94-223513,NJP11-073033,OE18-6891,CMS49-820,JOSAA30-1698,
PRL105-013901,APL98-043101,prl108-083902,PRA86-033825,OE21-10454,APL105-011109, JPCC118-30170,nanoscale6-9093}%
, and the concept of superscatterer has been extended to acoustics \cite%
{APA99-843,FP7-319}.

Superscatterer was first proposed in the framework of transformation
optics, where both electric and magnetic anisotropic inhomogeneous materials
were needed \cite{OE16-18545}. In order to loose the requirement, Ruan and
Fan \cite{PRL105-013901,APL98-043101} proposed a kind of subwavelength
superscatterer by engineering an overlap of resonances of different
plasmonic modes where they only use multilayered isotropic plasmonic
structures. To extend their superscatterer into the deep-subwavelength
scale, we put forward the idea of using graphene monolayer to realize
cylindrical superscatterers with the advantage of active tunability \cite%
{OL40-1651}. Since graphene is a two dimensional carbon sheet with only one
atom thick \cite{RMP,science332-1291}, this superscatterer can be treated as
one kind of ``surface superscatterers''.
However, in practical applications, three dimensional (3D)
superscatterers are more promising since the objects to be
superscattered are usually 3D.
Different from the two dimensional (2D) cylindrical case,
the parallel polarized and vertically polarized fields cannot been
decoupled from the incident electromagnetic field in the 3D case.
Thus, it is desirable to extend the
notion of ``superscattering by a surface" to the 3D case.

Furthermore, similarities between quantum physics and wave optics have been hot
topics in recent decades \cite{bookQCA,LPR3-243}. On one hand, optics
provides a fertile ground to simulate some basic concepts in quantum
mechanics and quantum field theory, such as PT symmetry \cite%
{prl100-103904,nphys6-192,prl103-093902}, supersymmetry \cite%
{prl110-233902,optica}, Anderson localization \cite{nphoton7-197}, and
quantum walks \cite{bookQRW}. On the other hand, methods from quantum
mechanics such as path integral method has been applied to study light
propagation in inhomogeneous materials \cite{bookQTFIGIW}, and ideas in
solid state physics and atomic physics can be transferred to optics to design some
practical devices, such as photonic crystals \cite{bookPC} and photonic
lattices \cite{PR518-1}. In Ref. \cite{arxiv}, Liu et. al. gave a geometric
interpretation for resonances of plasmonic nanoparticles by applying the
Bohr quantization condition in quantum mechanics to optics, which can
explain directly the scattering features of plasmonic nanoparticles.
Specially, we will use this method to interpret our superscattering
phenomenon.

In this paper, we theoretically propose that atomically thin
3D shell-shaped superscatterers can be designed by graphene monolayers. The
first-order resonance of transverse magnetic (TM) graphene plasmons is excited to enhance
the scattering cross section of the bare subwavelength dielectric particle.
Applying Bohr model to optics, the superscattering phenomenon can be
intuitively understood and interpreted. Moreover, different from those for
the TM case, we find that superscattering is hard to occur by exciting the
resonance of transverse electric (TE) graphene plasmons due to their poor field confinements.

\section{Structure and model}
The structure of our spherical shell-shaped superscatterer is shown in Fig. \ref%
{superscatterer} (a), where a graphene monolayer is coated on the dielectric
particle of radius $R$, and a $x$-polarized plane wave is incident from air
onto the structure. The dielectric particle is the object to be superscattered.
The incident electric field is $\mathbf{E}%
=E_{0}e^{ik_{0}z}\hat{x}$ with the time dependence of $\exp \left( -i\omega
t\right) $, where $k_{0}=\omega \sqrt{\varepsilon _{0}\mu _{0}}$ is the
wave number in free space and $\omega $ is the angular frequency of the
incident field. The relative permittivity of the dielectric particle is $%
\varepsilon _{r}$, and the relative permeability is $\mu _{r}=1$.
This kind of structure can be realized experimentally by using wet transfer
technique similar with fabrication of graphene/GaAs heterostructure \cite{ne16-310}
and graphene can be well grown by CVD reaction between CH$_{\text{4}}$ and H$_{\text{2}}$
on cooper substrate, which can be etched away using acid solution. The suspending
graphene can be transferred to arbitrary substrate subsequently. Previously, we also have
realized graphene surrounded ZnO nanowire structure, demonstrating the feasibility of
graphene coated dielectric particle experimentally \cite{oe20-A706}.

\begin{figure}[tbp]
\vspace{0.0cm} \includegraphics[width=6.0cm]{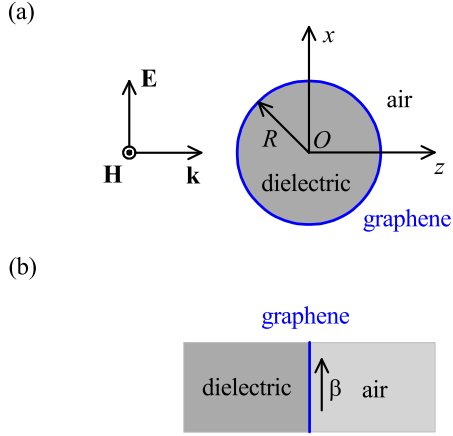} \vspace{0.3cm}
\caption{(a) Cross-sectional view of the spherical shell-shaped superscatterer at $y=0$
plane, where the dielectric particle (dark grey area) to be superscattered
is coated with the
graphene monolayer (blue area). A $x$-polarized plane wave is incident from
air onto the superscatterer. (b) Structure of the equivalent planar
waveguide, where the graphene monolayer (blue area) is separated by a
semi-infinite dielectric medium (dark gray area) and the air (light gray
area). The graphene plasmons propagate along the graphene surface in the
direction indicated by an arrow.}
\label{superscatterer}
\end{figure}

Analytical Mie scattering theory is applied to solve this scattering problem
\cite{bookEWPRS,bookASLSP}. Different from the 2D case where
TM and TE waves are defined according to the axis of the cylinder \cite{OL40-1651},
the electromagnetic fields in 3D case can be decomposed
into TM and TE waves with respect to the radial direction $\hat{r}$, namely $\mathbf{H}_{\text{TM}%
}=\nabla \times \left( \Phi _{\text{TM}}\mathbf{r}\right) $ and $\mathbf{E}_{%
\text{TE}}=\nabla \times \left( \Phi _{\text{TE}}\mathbf{r}\right) $, where $%
\Phi _{\text{TM}}$ is the electric potential and $\Phi _{\text{TE}}$ is the
magnetic potential. For the incident field, its corresponding electric
potential $\Phi _{\text{TM}}$ and magnetic potential $\Phi _{\text{TE}}$ are%
\begin{equation}
\Phi _{\text{TM}}^{i}=-\frac{E_{0}\cos \phi }{k_{0}\eta _{0}r}%
\sum_{n=1}^{\infty }a_{n}\psi _{n}\left( k_{0}r\right) P_{n}^{1}\left( \cos
\theta \right)  \label{phi_TM_i}
\end{equation}
and%
\begin{equation}
\Phi _{\text{TE}}^{i}=\frac{E_{0}\sin \phi }{k_{0}r}\sum_{n=1}^{\infty
}a_{n}\psi _{n}\left( k_{0}r\right) P_{n}^{1}\left( \cos \theta \right)\,,
\label{phi_TE_i}
\end{equation}
respectively, where $a_{n}=\left( -i\right) ^{-n}\left( 2n+1\right) /n\left(
n+1\right) $ is the expansion coefficient. Similarly, scalar potentials are%
\begin{equation}
\Phi _{\text{TM}}^{s}=-\frac{E_{0}\cos \phi }{k_{0}\eta _{0}r}%
\sum_{n=1}^{\infty }a_{n}s_{n}^{\text{TM}}\xi _{n}\left( k_{0}r\right)
P_{n}^{1}\left( \cos \theta \right)  \label{phi_TM_s}
\end{equation}
and%
\begin{equation}
\Phi _{\text{TE}}^{s}=\frac{E_{0}\sin \phi }{k_{0}r}\sum_{n=1}^{\infty
}a_{n}s_{n}^{\text{TE}}\xi _{n}\left( k_{0}r\right) P_{n}^{1}\left( \cos
\theta \right)  \label{phi_TE_s}
\end{equation}
for the scattered electromagnetic field at $r>R$, and
\begin{equation}
\Phi _{\text{TM}}^{in}=-\frac{E_{0}\cos \phi }{k\eta r}\sum_{n=1}^{\infty
}a_{n}b_{n}^{\text{TM}}\psi _{n}\left( kr\right) P_{n}^{1}\left( \cos \theta
\right)  \label{phi_TE_int}
\end{equation}
and%
\begin{equation}
\Phi _{\text{TE}}^{in}=\frac{E_{0}\sin \phi }{kr}\sum_{n=1}^{\infty
}a_{n}b_{n}^{\text{TE}}\psi _{n}\left( kr\right) P_{n}^{1}\left( \cos \theta
\right)  \label{phi_TE_int}
\end{equation}
for the electromagnetic field inside the dielectric particle ($r\leq R$),
where $k=k_{0}\sqrt{\varepsilon _{r}\mu _{r}}$ is the wavenumber in the
dielectric particle, $s_{n}^{\text{TM}}$, $s_{n}^{\text{TE}}$, $b_{n}^{\text{%
TM}}$, and $b_{n}^{\text{TE}}$ are unknown expansion coefficients, $\psi
_{n}\left( k_{0}r\right) $ and $\xi _{n}\left( k_{0}r\right) $ are the $n$%
-th order Riccati--Bessel function of the first kind and the third kind,
respectively \cite{bookHMFFGMT}. Meanwhile, $\eta =\sqrt{\mu _{0}\mu
_{r}/\varepsilon _{0}\varepsilon _{r}}$ and $\eta _{0}=\sqrt{\mu
_{0}/\varepsilon _{0}}$ are the impedances of the dielectric particle and
free space, respectively. Since the thickness of graphene is extremely small
compared with the radius of the spherical dielectric particle, the graphene coating can
be well characterized as a two dimensional homogenized conducting film with surface
conductivity $\sigma _{g}$, where the non-spherical elements and microscopic details
are neglected \cite{science332-1291,JAP103-064302,PRB91-125414}.
According to the continuity conditions at $r=R$, the scattering
coefficient can be obtained as%
\begin{eqnarray}
&&s_{n}^{\text{TM}} =\frac{-\psi _{n}^{\prime }\left( k_{0}R\right) t_{n}^{%
\text{TM}}+\left( \eta /\eta _{0}\right) \psi _{n}\left( k_{0}R\right) \psi
_{n}^{\prime }\left( kR\right) }{\xi _{n}^{\prime }\left( k_{0}R\right)
t_{n}^{\text{TM}}-\left( \eta /\eta _{0}\right) \xi _{n}\left( k_{0}R\right)
\psi _{n}^{\prime }\left( kR\right) }\,,  \label{1} \\
&&s_{n}^{\text{TE}} =\frac{-\psi _{n}\left( k_{0}R\right) t_{n}^{\text{TE}%
}+\left( \eta /\eta _{0}\right) \psi _{n}^{\prime }\left( k_{0}R\right) \psi
_{n}\left( kR\right) }{\xi _{n}\left( k_{0}R\right) t_{n}^{\text{TE}}-\left(
\eta /\eta _{0}\right) \xi _{n}^{\prime }\left( k_{0}R\right) \psi
_{n}\left( kR\right) }\,,  \label{SN_TM_TE}
\end{eqnarray}%
where $t_{n}^{\text{TM}}=\psi _{n}\left( kR\right) +i\sigma _{g}\eta \psi
_{n}^{\prime }\left( kR\right) $, $t_{n}^{\text{TE}}=\psi _{n}^{\prime
}\left( kR\right) -i\sigma _{g}\eta \psi _{n}\left( kR\right) $, and $\sigma
_{g} $ is the surface conductivity of graphene. We define the normalized
scattering cross section (NSCS) as
\begin{equation}
\mathrm{NSCS}=\sum\limits_{n=1}^{\infty }\left( 2n+1\right) \left(
\left\vert s_{n}^{\text{TM}}\right\vert ^{2}+\left\vert s_{n}^{\text{TE}%
}\right\vert ^{2}\right)\,,  \label{NSCS}
\end{equation}
which is normalized by $\lambda ^{2}/2\pi $. Note that when $\sigma _{g}=0$, it
reduces to the case of scattering by a bare spherical dielectric particle.

The aforementioned scattering model is related to its equivalent one dimensional
planar waveguide by Bohr model, where the structure of the planar waveguide
is shown in Fig. \ref{superscatterer} (b). To support a resonance, Bohr
condition requires that the phase accumulation along an enclosed optical
path should be an integral number of $2\pi$, namely
\begin{equation}
\oint \beta dl=2n\pi\,,  \label{Bohr_condition}
\end{equation}%
where $n$ is the order of the resonance, $\beta$ is the corresponding
propagation constant in the equivalent planar waveguide, and the integral is
calculated along the circumference of the sphere since graphene plasmons
propagate as a surface wave along the graphene surface with evanescent
fields in the perpendicular direction.

Meanwhile, in order to get the NSCS of the superscatterer shown in Fig. \ref%
{superscatterer} (a) from Mie scattering theory and the propagation constant
of the equivalent planar waveguide shown in Fig. \ref{superscatterer} (b), the
surface conductivity of graphene is calculated according to the Kubo formula
\begin{equation}
\sigma_{g}\left( \omega ,\mu _{c},\Gamma ,T\right) =\sigma _{\text{intra}}+\sigma
_{\text{inter}}\,,  \label{sigma_g}
\end{equation}
where
\begin{equation}
\sigma _{\text{intra}}=\frac{ie^{2}k_{B}T}{\pi \hbar ^{2}\left( \omega +i2\Gamma
\right) }\left[ \frac{\mu _{c}}{k_{B}T}+2\ln \left( e^{-\mu
_{c}/k_{B}T}+1\right) \right]  \label{sigma_intra}
\end{equation}%
is due to intraband contribution, and
\begin{equation}
\sigma _{\text{inter}}=\frac{ie^{2}\left( \omega +i2\Gamma \right) }{\pi \hbar ^{2}}%
\int_{0}^{\infty }\frac{f_{d}\left( -\varepsilon \right) -f_{d}\left(
\varepsilon \right) }{\left( \omega +i2\Gamma \right) ^{2}-4\left(
\varepsilon /\hbar \right) ^{2}}d\varepsilon  \label{sigma_inter}
\end{equation}%
is due to interband contribution \cite{JAP103-064302,JP19-026222}. In the
above formula, $-e$ is the charge of an electron, $\hbar =h/2\pi $ is the
reduced Plank's constant, $\Gamma $ is the phenomenological scattering rate
that is assumed to be independent of the energy $\varepsilon $, $f_{d}\left(
\varepsilon \right) =1/\left[ e^{\left( \varepsilon -\mu _{c}\right)
/k_{B}T}+1\right] $ is the Fermi-Dirac distribution, $k_{B}$ is the
Boltzmann's constant, $T$ is the temperature, and $\mu _{c}$ is the chemical
potential which can be tuned by a gate voltage and/or chemical doping. In
the following, we choose $\Gamma =0.11$ meV and $T=300$ K \cite{JAP103-064302}.

\section{Results and discussion}
\begin{figure}[tbp]
\centering
\vspace{-0.0cm} \centerline{\includegraphics[width=8.5cm]{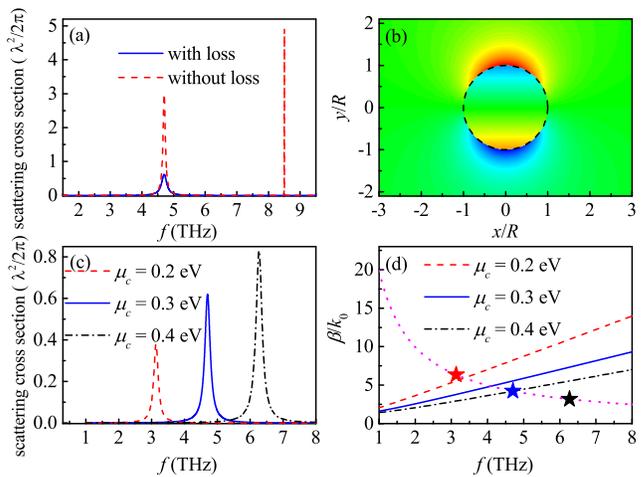}} \vspace{%
-0.0cm}
\caption{(a) The dependence of the normalized scattering cross section
(NSCS) on the incident frequency $f$, where the graphene monolayer
with $\mu_{c}=0.3$ eV is
considered to be lossless or lossy. (b) Normalized $H_{\protect\theta}$
field distribution at $z=0$ plane with $f=4.7$ THz when the optical loss of
graphene is involved. (c) NSCSs for different frequencies when $\protect\mu%
_c=0.2$ eV, $0.3$ eV, and $0.4$ eV, respectively. (d) Dispersion relations
of TM graphene plasmons in the equivalent planar waveguide for $\protect\mu%
_c=0.2$ eV, $0.3$ eV, and $0.4$ eV, respectively. The dotted magenta curve denotes
the Bohr condition for superscattering, and three stars in the curve correspond to the
three resonate frequencies in (c). The other parameters are $\protect%
\varepsilon _{r}=1.44$ and $R=0.24$ $\protect\mu $m.}
\label{superscattering_TM}
\end{figure}

To demonstrate the superscattering phenomenon for subwavelength particles,
we let $\varepsilon _{r}=1.44$, $R=0.24$ $\mu $m,
and $\mu _{c}=0.3$ eV \cite{JAP103-064302}.
Under these parameters, Fig. \ref{superscattering_TM} (a) schematically
shows the dependence of the NSCS on the incident frequency $f$, where the
graphene monolayer is considered to be lossless or lossy. When the optical
loss of graphene is neglected with $\Gamma =0$ meV,
two sharp resonances occur at $f=\omega /2\pi
=4.7$ THz and $f=8.5$ THz, respectively. The corresponding NSCSs are close
to their single channel limits $2n+1$ \cite{PRL105-013901,PRL97-263902},
where the resonances at $f=4.7$ THz and $f=8.5$ THz are dominated by the $n=1
$ and $n=2$ scattering terms, respectively. When the optical loss is
involved with $\Gamma \ne 0$ meV \cite{nnano6-630,SSC,nphoton7-394},
the NSCSs at both two resonances
decrease from their corresponding single channel limits
due to the damping effects of graphene plasmons.
As shown in Fig. \ref{superscattering_TM} (a), the NSCS at the first order
resonance ($n=1$) decreases to 0.62, while the NSCS at
the second order resonance ($n=2$) decreases more sharply and is equal to
zero approximately.
This is because the second order resonance is more susceptible to the
loss \cite{PRL105-013901}.
As shown in Fig. \ref{superscattering_TM} (b), the normalized $H_{\theta }$
field distribution exhibits a first order resonant mode at $f=4.7$ THz.
Since the chemical potential of graphene can be tuned by a gate voltage
and/or chemical doping \cite{JAP103-064302}, Fig. \ref{superscattering_TM}
(c) shows NSCSs for different frequencies at different values of chemical
potential, where the decrease of peak NSCS at lower resonate frequencies is
caused by the increasing optical loss of graphene \cite{JAP103-064302}. The NSCS
of the bare dielectric particle is in the order of $10^{-6}$, this indicates
that coating with a properly doped and/or gated graphene monolayer can
greatly enhance the scattering by five orders of magnitude at certain
frequencies. Meanwhile, larger chemical potential leads to superscattering
at higher frequency, which can be understood with Bohr model.

In the above frequency range, the surface conductivity of graphene has a
positive imaginary part, which implies that TM graphene plasmons are supported with
the components $H_\theta$, $E_\phi$, and $E_r$ \cite{science332-1291,JAP103-064302}.
For the spherical shell-shaped superscatterer, the dispersion
relation of TM graphene plasmons in the equivalent planar waveguide is
\begin{equation}
\frac{\varepsilon _{r}}{k_{1}}+\frac{1}{k_{2}}+i\frac{\sigma _{g}}{\omega
\varepsilon _{0}}=0,  \label{dispersion_TM}
\end{equation}
where $k_{1}=\left( \beta ^{2}-k_{0}^{2}\varepsilon _{r}\right) ^{1/2}$, $%
k_{2}=\left( \beta ^{2}-k_{0}^{2}\right) ^{1/2}$, and $\beta $ is the
propagation constant.
Meanwhile, we define $\delta_d=1/k_1$ and $%
\delta_a=1/k_2$ as the penetration depths in the dielectric medium and air,
respectively. Fig. \ref{superscattering_TM} (d) shows the dispersion
relations for $\mu _{c}=0.2$ eV, $0.3$ eV, and $0.4$ eV, where the optical
loss of graphene is neglected. The first order Bohr condition $\beta =1/R$
for superscattering is denoted by the dotted magenta curve.
According to the Bohr model, superscattering occurs at the
intersection points of dispersion relations and the Bohr condition, which
exhibits blue shift when the chemical potential of graphene increases.
For comparison, resonant
frequencies for $\mu _{c}=0.2$ eV, $0.3$ eV, and $0.4$ eV are calculated from
Mie scattering theory and marked by three stars
in the dotted magenta curve. Clearly, the stars exhibit blue shift similarly.

\begin{figure}[tbp]
\vspace{0.0cm} \centering
\centerline{\includegraphics[width=8.5cm]{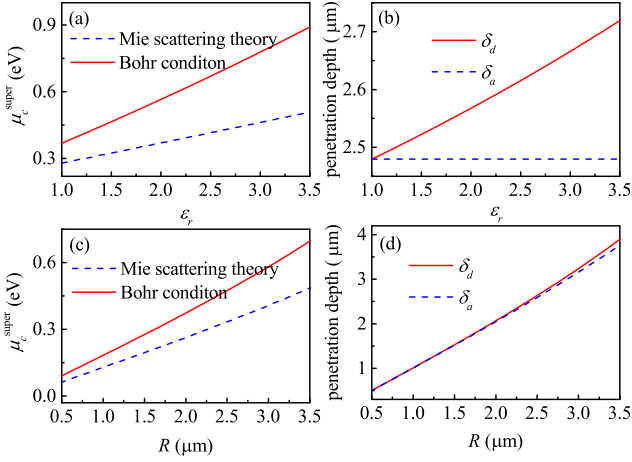}} \vspace{-0.0cm}
\caption{Left panel: The dependences of $\protect\mu _{c}^{\text{super}}$
on the (a) relative permittivity $\protect\varepsilon _{r}$ and (c) radius $R$,
respectively, where superscattering occurs when $\protect\mu _{c}=\protect%
\mu _{c}^{\text{super}}$ under the corresponding value of $\protect%
\varepsilon _{r}$ or $R$. The solid red curve and the dashed blue curve are
calculated by Bohr condition and Mie scattering theory, respectively. Right
panel: The dependences of the penetration depth ($\protect\delta _{d}$ for
the dielectric medium and $\protect\delta _{a}$ for the air) on the (b)
relative permittivity $\protect\varepsilon _{r}$ and (d) radius $R$,
respectively. The parameters are $R=0.24$ $\protect\mu $m for (a) and (b), $%
\protect\varepsilon _{r}=1.44$ for (c) and (d), and $f=5$ THz for (a)-(d).}
\label{figure3}
\end{figure}

As shown in Fig. \ref{superscattering_TM} (d), star deviates from the
intersection point at higher frequencies. In order to interpret this
deviation, we first analyse a simple case. Fig. \ref{figure3} (a)
shows the dependence of the chemical potential $\mu _{c}^{\text{super}}$
on the relative permittivity of dielectric particle, where
$\mu _{c}^{\text{super}}$ is the value of chemical potential when superscattering
occurs. Once the radius of the sphere is fixed, the propagation constant $%
\beta $ is determined by the Bohr condition. However, the increase of the
relative permittivity $\varepsilon _{r}$ leads to the increase of propagation constant $\beta $,
which further leads to the increase of the chemical potential to ensure the invariance
of the propagation constant. Moreover, since the penetration depth of
graphene plasmons in the dielectric medium $\delta _{d}$ becomes large when $%
\varepsilon _{r}$ increases as shown in Fig. \ref{figure3} (b), the calculation of the
path integral along the circumference of the sphere is not accurate any more \cite{arxiv}.
This produces errors to the calculation of
propagation constant $\beta $ and chemical potential $\mu _{c}$. Actually,
the growth rate of chemical potential slows down to match a shorter
integration path.

Although the radius of the sphere $R$ is not related directly to
the penetration depths of graphene plasmons, a small value of propagation
constant $\beta $ is required to support the resonance in a large sphere.
This leads to the increase of chemical potential $\mu _{c}^{\text{super}}$
and penetration depths $\delta _{d}$ and $\delta _{a}$, which can be seen
clearly from Fig. \ref{figure3} (c)-(d). Since the penetration depth in the
dielectric medium is larger than that in the air, the growth rate of
chemical potential slows down to match a shorter integration path.
Then we can use the same method to interpret the deviation in Fig. \ref%
{superscattering_TM} (d). The increase of incident frequency is equivalent
to an increase of the radius, thus deviation between Mie scattering theory
and Bohr condition becomes large at high frequencies.

\begin{figure}[tbp]
\centering
\vspace{0.1cm} \centerline{\includegraphics[width=6cm]{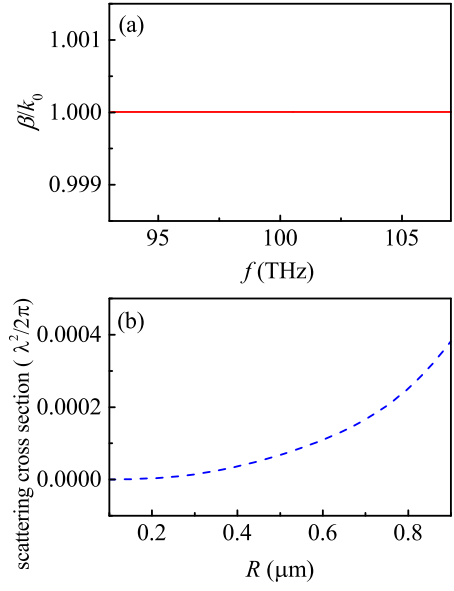}}
\vspace{-0cm}
\caption{(a) Dispersion relation of TE graphene plasmons. (b) The
dependence of the normalized scattering cross section (NSCS) on the radius
$R$ with $f=100$ THz. The parameters are $\protect\mu_{c}=0.2$ eV and $%
\protect\varepsilon_{r}=1$.}
\label{figure4}
\end{figure}

From above analysis, we know that Bohr model provides an intuitive way to
understand the superscattering of subwavelength particles and to design
corresponding 3D superscatterers. For example, based on the
Bohr model, we can predict the nonexistence of superscattering via the
resonance of TE graphene plasmons. We let $f=100$ THz, $\varepsilon _{r}=1$,
and $\mu _{c}=0.2$ eV, since TE graphene plasmons do not exist for $%
\varepsilon _{r}=1.44$. Under these parameters, the surface conductivity of
graphene has a negative imaginary part, which implies that
the TE graphene plasmons are supported with the components
$E_\theta$, $H_\phi$, and $H_r$ \cite{science332-1291,JAP103-064302}.
The dispersion relation of TE
graphene plasmons in the equivalent planar waveguide is
\begin{equation}
k_{1}+k_{2}-i\sigma _{g}\omega \mu _{0}=0,  \label{dispersion_TE}
\end{equation}%
where $k_{1}$, $k_{2}$, and $\beta $ are the same as in Eq. (\ref%
{dispersion_TM}).
According to Bohr condition, superscattering should occur
at $R=1/k_{0}=3/2\pi $ $\mu $m, where we have replaced the propagation
constant of TE graphene plasmons by the wave number in air since $\beta
\approx k_{0}$ as shown in Fig. \ref{figure4} (a). However, no
superscattering occurs as shown in Fig. \ref{figure4} (b). The
reason is that no appropriate integration path can be used to match the Bohr
condition due to poor field confinements of TE graphene plasmons.

\section{Conclusions}
In conclusion, we demonstrate for the first time that graphene monolayers can be used for
atomically thin spherical shell-shaped superscatterer designs. The first-order resonance
of TM graphene plasmons can be excited to enhance the scattering cross
section of the bare subwavelength dielectric particle by five orders of
magnitude. Scientifically, Bohr model is used to understand and interpret the
superscattering phenomenon. Based on the analysis of Bohr model, we show that
contrary to the TM case,
superscattering is hard to occur by exciting the resonance of TE graphene
plasmons due to their poor field confinements. Our work will provide
theoretical guidance for the future experimental design of superscatterers
and other graphene based devices, which have great potential applications
in sub-wavelength plasmonics.

\section*{Acknowledgement}
We are grateful to Mr. Hamza Madni for revising the manuscript. This work was sponsored by the
National Natural Science Foundation of China
under Grants No. 61322501, No. 61574127, and No. 61275183, the National Program for Special
Support of Top-Notch Young Professionals, the Program for New Century Excellent
Talents (NCET-12-0489) in University, the Fundamental Research Funds for the Central
Universities, and the Innovation Joint Research Center for Cyber-Physical-Society System.





\end{document}